# Novel Numerical Methods for Measuring Distributions of Space Charge and Electric Field in Solid Dielectrics with Deconvolution Algorithm[*]

Luoquan HU, Yewen ZHANG, Feihu ZHENG
Pohl Institute of Solid State Physics, Tongji University, Shanghai 200092, China

**Abstract**
The deconvolution algorithm for measuring distribution of space charge under dc by the pressure wave propagation (PWP) method is studied in this paper. A new Fredholm integral equation of first kind, including a space charge distribution without a partial differential operator is presented. Numerical methods based on Tikhonov regularization for solving this integral equation and the original PWP equation are studied. Numerical simulation is studied for the effect of signal-to-noise ratio (SNR), and comparison with other algorithms is discussed. The numerical solution of an electric field distribution from measurements of a LDPE specimen is obtained successfully.

## 1 Introduction

Space charge in solid dielectrics plays a very important role for the properties of insulation polymer under high voltage [1, 2]. Since 1980's，many researchers worked on the space charge distribution measurement for solid dielectrics, and studied the behavior of space charge in it. PWP method is an important method measuring space charge distribution in solid dielectrics. This method has been widely used in space charge research for understanding the charge behavior in solid dielectrics. However when using PWP method measurement, the electric field $E(z)$ in a specimen is included in an integral equation, named Fredholm Integral Equations of the First Kind (FIEFK) [3]. In order to get the real electric field distribution, deconvolution algorithm should be utilized. However, it is relatively difficult due to an ill-posed problem in FIEFK. There are some deconvolution methods to solve FIEFK [4, 5, 6], which can be generally divided generally into two kinds, one is Tikhonov regularization [7, 8], the other is iterative regularization [9, 10]. M. Abou-Dakka tried modified inverse matrix method to treat with high SNR data, but the result of this method is not steady [11]. If the voltage of open circuit is measured, electric field $E(z)$ is included in a standard FIEFK. From this equation a direct deconvolution method can be adopted. T. Ditchi has studied this problem in detail by means of the way introduced by Philips [12, 13, 14]. However, the results are not perfect. If the short circuit current is measured, differential form of pressure wave is included in a standard FIEFK. Since differential form of pressure wave only can be determined appropriately, this equation is more difficult for direct deconvolution method, unless some special methods are adopted. This paper introduces a new FIEFK through the mathematical transformation of original integral equation. With Tikhonov regularization, distributions of space charge and electric field in solid dielectrics are studied for a PE sample.

## 2 Theoretical Analysis
### 2.1 FIEFK including distribution of space charge $\rho(z)$

Figure 1 shows the measurement principle schema of PWP method [15]. The analysis assumes that a pressure wave $p(z,t)$ propagates along $-z$ direction at the velocity $u$, $z_f$ is the front position of the pressure wave at time $t$, $a$ and $b$ are original positions of electrodes before the pressure wave arrives at specimen. In this case, the short circuit current can be shown as [3]:

$$I(t) = \chi C_0 G(\varepsilon) \int_0^{z_f} E(z,t) \frac{\partial}{\partial t} p(z,t) dz \qquad (1)$$

where $I(t)$ is the current signal of measurement, $\chi$ is the ratio of compression of materials,

---

$C_0$ is the initial capacitance of specimen, $G(\varepsilon)$ is a constant relating to polarization characteristics of materials, $E(z,t)$ is the electric field in the specimen; $p(z,t)$ is the amplitude of pressure wave at location $z$ and time $t$.

For a planar specimen, according to the basic principle of the PWP method, considering the effect of attenuation and dispersion, the pressure wave profile at any position $z$ within the bulk can be calculated by the inverse Fourier transform based on the following equation [12]:

$$p(z, t) = \mathbb{F}^{-1}\left[\mathbb{F}(p(0, t)) \cdot [A(\upsilon)]^z\right]$$
$$= \mathbb{F}^{-1}\left[\mathbb{F}(p(0, t)) \cdot [A(\upsilon)]^z\right] \quad (2)$$

where $A(\upsilon) = \left\{\dfrac{\mathbb{F}[p(d_0,t)]}{\mathbb{F}[p(0,t)]}\right\}^{-d_0}$ is a degradation factor, $d_0$ is the thickness of the sample, $z$ is the abscissa of the pressure wave, $\mathbb{F}$ and $\mathbb{F}^{-1}$ represent Fourier transform and inverse Fourier transform, respectively.

Let $A = u\chi C_0 G(\varepsilon)$, $u = -\dfrac{dz}{dt}$, and integrating equation (1) by parts, assuming that the pressure is a continuous function at $z_f$, therefore $p(z_f, t) = 0$, then:

$$I(t) = \chi C_0 G(\varepsilon) \int_0^{z_f} E(z,t) \dfrac{\partial}{\partial z} p(z,t) \dfrac{dz}{dt} dz,$$
$$I(t) = -A \int_0^{z_f} E(z,t) \dfrac{\partial}{\partial z} p(z,t) dz, \text{ then:}$$
$$I(t) - AE(0,t)p(0,t) = A \int_0^{z_f} \dfrac{\partial E(z,t)}{\partial z} p(z,t) dz \quad (3)$$

where $p(0,t)$ is the pressure at the interface of the specimen, $E(0,t)$ is electric field at the interface of an electrode at $z=0$, which will be discussed in the next section. Using Poisson's equation, supposed that $E(0,t)$ and $\rho(z,t)$ are not dependent on $t$ (It is reasonable the time of every measurement is so short that the distributions of electric field and space charge do not change a lot before and after penetration of the pressure wave, therefore $E(0,t)$ and $\rho(z,t)$ can be approximately treated as constant during the measurement process), equation (3) can be changed into:

$$I(t) - AE(0)p(0,t) = \dfrac{A}{\varepsilon_0 \varepsilon_r} \int_0^{z_f} \rho(z)p(z,t)dz \quad (4)$$

Equation (4) is a FIEFK including the distribution of space charge $\rho(z)$. With approximation of finite dimensions, it can be written as the following:

$$I(i) - AE(0)p(0,i) = \dfrac{A}{\varepsilon_0 \varepsilon_r} \sum_{j=1}^{n} p(i,j)\rho(j)\Delta z \quad (5)$$

where $i$ is time index ($i=1, 2,\ldots m$), $j$ is location index ($j=1, 2,\ldots n$). For the sake of convenience of computing, often let $m=n$. Let $A' = A\Delta z$, equation (5) can be rewritten as the following:

$$I(i) - A'E(0)p(0,i) = \dfrac{A'}{\varepsilon_0 \varepsilon_r} \sum_{j=1}^{n} p(i,j)\rho(j) \quad (6)$$

**2.2 Deconvolution algorithm based on Tikhonov regularization**

With the introduction of the notations: $I = \varepsilon_0 \varepsilon_r [I(i) - A'E(0)p(0,i)]$, $D = \rho(j)$, $P = A'P[i,j]$, equation (6) can be changed into:

$$I = PD \quad (7)$$

where $I=I[n, 1]$, $D=D[n, 1]$, and $P=P[n, n]$, their dimensions are $n$; and $P$ is a $n \times n$ matrix, it is constructed from equation (2). With $P$ considered as an operator and the application of Tikhonov regularization technique [7, 8, 16], equation (7) can be changed into the following:

$$P^T I = \left[ P^T P + \alpha H \right] D \tag{8}$$

where $H$ is a special matrix, which is related to *a priori* knowledge, and can be replaced by a unit matrix; $\alpha$ is regularization factor [7], $P^T$ is accompanied matrix of $P$. Therefore

$$D = \left[ P^T P + \alpha H \right]^{-1} P^T I \tag{9}$$

### 2.3 Determination of $E(0)$ with iterative method

Define $E(0) = E(z)|_{z=0}$, the electric field on the interface of electrode at $z = 0$. If there is charge injection from electrodes, homo-charges can suppress the electric field in the domain near the electrodes, on the contrary, hetero-charges can increase the electric field in the domain near the electrodes.

The electric field inside the bulk (from 0 to $d_0$) can be obtained as the following

$$E(z) = \int_0^z \frac{\rho(\xi)}{\varepsilon_0 \varepsilon_r} d\xi + C, \quad 0 < z < d_0 \tag{10}$$

where $\rho(\xi)$ is the distribution of space charge, which can be calculated from (9) for a given $E(0)$; $C$ is an integral constant, which can be determined from the "equal electric potential" principle as the following:

$$V_a = \int_0^{d_0} E(z) \, dz \tag{11}$$

where $V_a$ is applied voltage. In order to get the exact $E(0)$, iterative method should be utilized. Iterative formula is the following:

$$E_{k+1} = g(E_k; \rho_k) \tag{12}$$

where $E_k$ is $E(0)$ of $k^{th}$ iterative solution, $g$ is a function relation, which is a combination of (9), (10) and (11), and $\rho_k$ is derived from $E_k$.

The iterative algorithm is as following:

Step 1. Let $k = 0$, and $E_0(0) = 0$, then from equation (9) $\rho_0(z)$ is obtained;

Step 2. From $\rho_0(z)$ and equation (10) and (11), $E_1(z)$ is obtained;

Step 3. From $E_1(z)$, $E_1(0)$ is obtained. If $|E_{k+1} - E_k| \leq tolerance$, iterative process finishes.

### 2.4 FIEFK including distribution of electric field and its numerical deconvolution algorithm

Equation (1) can be rewritten as the following:

$$I(t) = B \int_0^{z_f} E(z,t) \frac{\partial p(z,t)}{\partial t} dz \tag{13}$$

where $B = \chi C_0 G(\varepsilon)$. Let $Y = [I(t)]$, $G = E(z,t)$, $Q = \frac{\partial p(z,t)}{\partial t}$, $F = BQ$, after discretization treatment, and let $Y = Y[n,1]$, $G = G[n,1]$, $F = F[n,n]$, then (13) can be changed as the following:

$$Y = GF \tag{14}$$

Using the Tikhonov technology introduced in section 2.2, the deconvolution solution of (14) can be obtained as the following:

$$G = [F^T F + \alpha I]^{-1} F^T Y \tag{15}$$

In practice, because $F$ is a differential matrix, its elements are accompanied with lots of noise, some denoise technology should be used. In the paper, wavelet packet denoise method is utilized.

# 3 Results and discussion
## 3.1 Numerical simulation from new FIEFK
### 3.1.1 Effects of SNR for numerical solutions

Pressure wave propagation matrix $P[i, j]$ is shown as figure 2, which was constructed from the experimental data for a LDPE specimen without space charge according to equation (2). The assumed distribution of space charge is shown in figure 3. Suppose that the velocity of pressure wave is 2000m/s, after the interaction with $P[i,j]$, short circuit current $I(t)$ is obtained, which is shown in the figure 3. The solutions of $\rho(z)$ for different SNR (signal-noise ratio) are shown in figure 4. In consequence, the solutions of $E(z)$ for different SNR are shown in figure 5. Where the true solution in figure 4, assumed space charge given in figure 3, is theoretical result strictly satisfied with equation (7). The true solution in figure 5, theoretical result from the equation (10), is derived from the true solution of the distribution of space charge.

From figure 4, it is shown that the numerical solution of space charge distribution is affected by the SNR of given data, the higher SNR, the more precise solution. This is because that SNR affects the regularization factor $\alpha$. For Tikhonov regularization method, when $\alpha \to 0$, numerical solution converges at the true solution of problem. But for a given SNR, its allowable $\alpha$ is determined, the higher SNR, the smaller allowable $\alpha$ is expected. Therefore with the increase in SNR, the numerical solution becomes more precise.

The numerical solutions of electric field distribution are shown in figure 5. Compared with figure 4, it can be found that numerical solution of electric field distribution is slightly affected by the SNR, which is caused by the integral. Integral can weaken the details of the numerical solution. It can be concluded it is practical to seek a numerical solution of electric field distribution for the experimental data.

### 3.1.2 Comparison with the existed method

T. Ditchi used the following equation for deconvolution [12]:

$$V(t) = A \int_0^{z_f} E(z) p(z, t) dz \tag{16}$$

This is a standard FIEFK. Here $V(t)$ is the integral of $I(t)$ in equation (1), or the voltage of open circuit. Based on the same current $I(t)$ shown in figure 3, the numerical solutions of the distribution of electric field for different SNR from equation (16) are shown in figure 6. Consequently, the solutions of $\rho(z)$ for different SNR are shown in figure 7, which are derived from Poisson equation. SNR's effect for numerical solution is the same as discussed in section 3.1.1.

With the comparison between figure 5 and figure 6, it is clear that the numerical solutions of electric field distribution by two methods are similar except a slight up-down shift, which is due to the integral operation of $I(t)$. However, numerical solutions of space charge distribution are slightly different as shown in figure 4 and figure 7. It is because that $\rho(z)$ in figure 7 is derived from the differential operation from $E(z)$, which can produce larger errors. Thus precision of $\rho(z)$ should be degraded.

### 3.1.3 Numerical solution of $E(z)$ from FIEFK with a differential matrix of pressure wave

The differential matrix of pressure wave propagation $F[i, j]$ is shown as figure 8, which is reconstructed from pressure wave propagation matrix $P[i, j]$ after denoise treatment. For a signal of SNR=40dB based on the assumed short circuit current $I(t)$ in figure 3, its numerical solution of $E(z)$ derived from equation (15) is shown in figure 9. With the comparison between figure 9 and figure 5 (the line of SNR=40dB), it can be seen that numerical solutions of $E(z)$ are similar from two different methods. However, the method from differential matrix is more complex then the one introduced in section 3.1.1.

### 3.2 Electric field distribution from measurements of a LDPE specimen

A planar specimen of LDPE is used, 1mm thickness, and thermoplastic semi-conductor electrodes, 50mm diameter, -40kV DC is applied, PWP measurement system is adopted, digital oscillograph with sampling time 800ns.

The original experimental data, the short circuit currents, are shown in figure 10. The current is the conduction current measured from the sampling resistance of the outer circuit. It is also the response of the displacement current in the dielectrics. In figure 10, line 1 is the signal for calibration, which is required to be measured in very short time after applying voltage. Line 2 and line 3 are the measurement signals of space charge, and the latter is measured 19 minutes later after the measurement of the former. Because noises of original data are so huge for deconvolution algorithm that orthogonal wavelet packet technology is required. After denoise treatment with wavelet packet, 4-point average method is used. The results after pre-treatments are shown in figure 11.

The deconvolution algorithm based on Tikhonov regularization from equation (4) is used to get the electric field distribution for line 2 and line 3 in figure 11. The numerical solutions of electric field are shown in figure 12, where line 1 is the solution of line 2 in figure 11, and line 2 is the solution of line 3 in figure 11. From figure 12, it can be found that the distribution of electric field is relatively steeper near the electrodes, which is consistent with the physical model. The electric field strength at anode of line 1 is less than that of line 2, which is caused by homo-charge injection from electrode. Inside the bulk of specimen, it can be also found that with the movement of space charge, the maximum point of electric field moves from anode to cathode.

### 4 Conclusions

The applications of deconvolution algorithm for distribution of space charge in solid dielectrics from a new FIEFK and a FIEFK with a differential matrix are successful. Numerical simulation demonstrates that numerical solution is acceptable, but SNR plays a very important role for deconvolution. Deconvolution from the new FIEFK is more effective than the existing method. Electric field distribution in a LDPE specimen shows that the deconvolution method is suitable for the study of space charge.

### Acknowledgement

The project is supported by National Science Foundation of China (No. 50277026), by key program of Ministry of Education of China (No.02100), and by the 973 projects of Ministry of Science and Technology of China (No.2001CB610406).

**Figure captions**

Figure 1.  Scheme of measurement principle of PWP method.

Figure 2.  Profile of pressure wave propagation (along the negative direction of z axes).

Figure 3.  Assumed distribution of space charge and short circuit current.

Figure 4.  The solutions of $\rho(z)$ for different SNR.

Figure 5.  The solutions of $E(z)$ for different SNR.

Figure 6.  The numerical solutions of $E(z)$ for different SNR using existing method.

Figure 7.  The numerical solutions of $\rho(z)$ for different SNR using existing method.

Figure 8.  Differential of profile of pressure wave propagation in figure 2

Figure 9.  The numerical solution of $E(z)$ for SNR=40dB using differential matrix method.

Figure 10. Original experimental data of LDPE specimen. Line 1 is the signal measured immediately after voltage application, and line 2 is measured after 37 minutes of voltage application.

Figure 11. Experimental data after pre-treatments. Line 1 and line 2 are corresponding to the curves in figure 10, respectively.

Figure 12. Numerical solutions of electric field distribution for line 2 in figure 11.

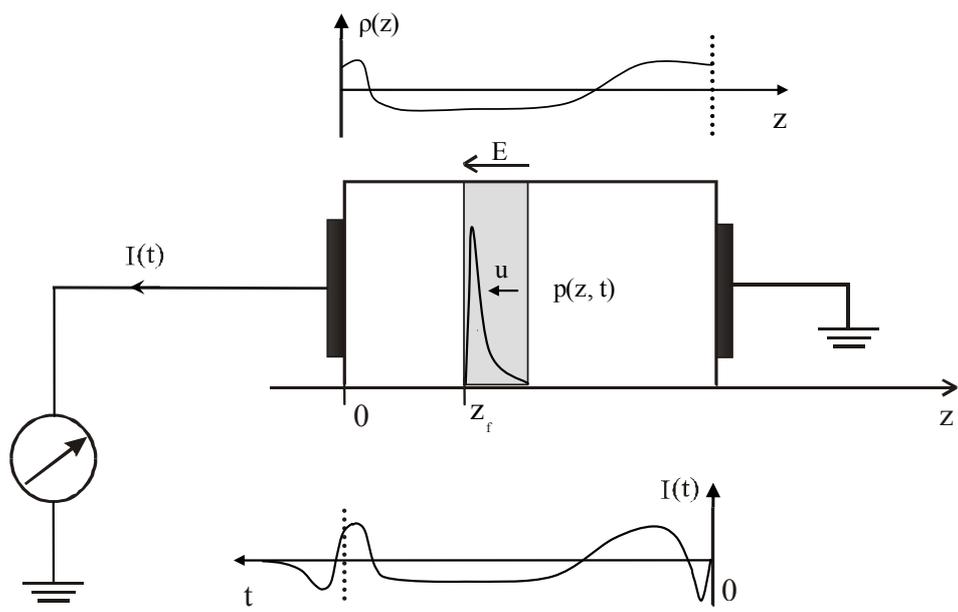

Figure 1. Luoquan HU *et al*

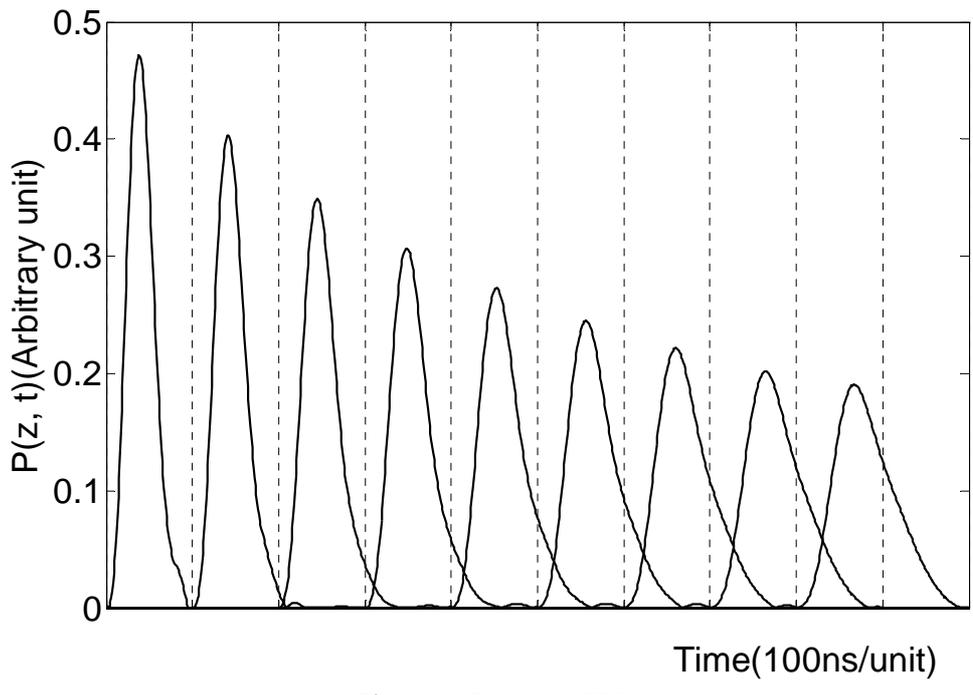

Figure 2. Luoquan HU *et al*

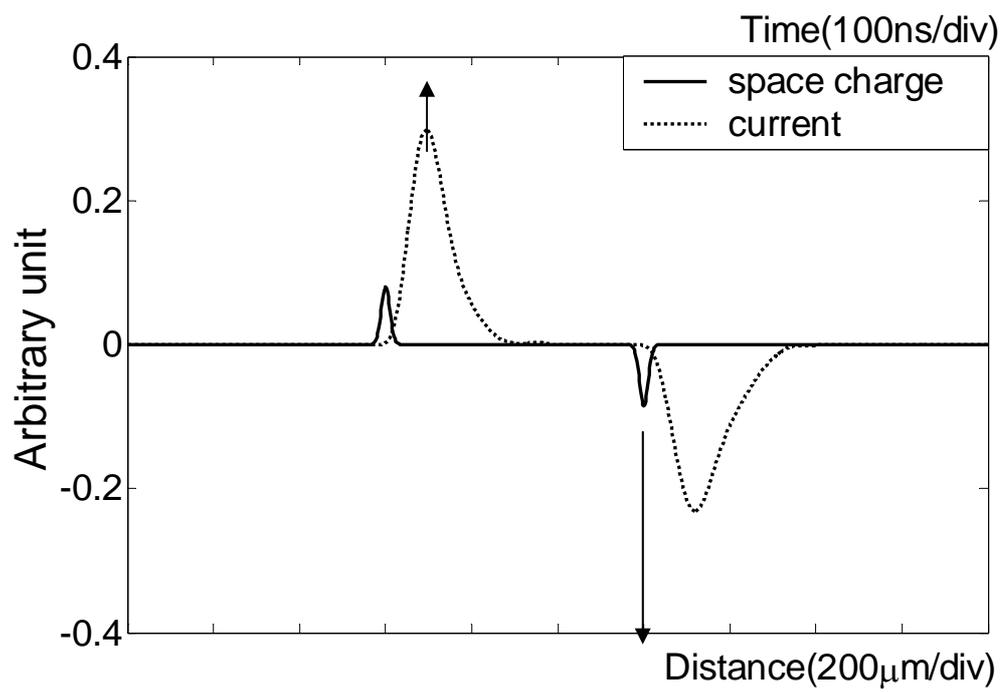

Figure 3. Luoquan HU *et al*

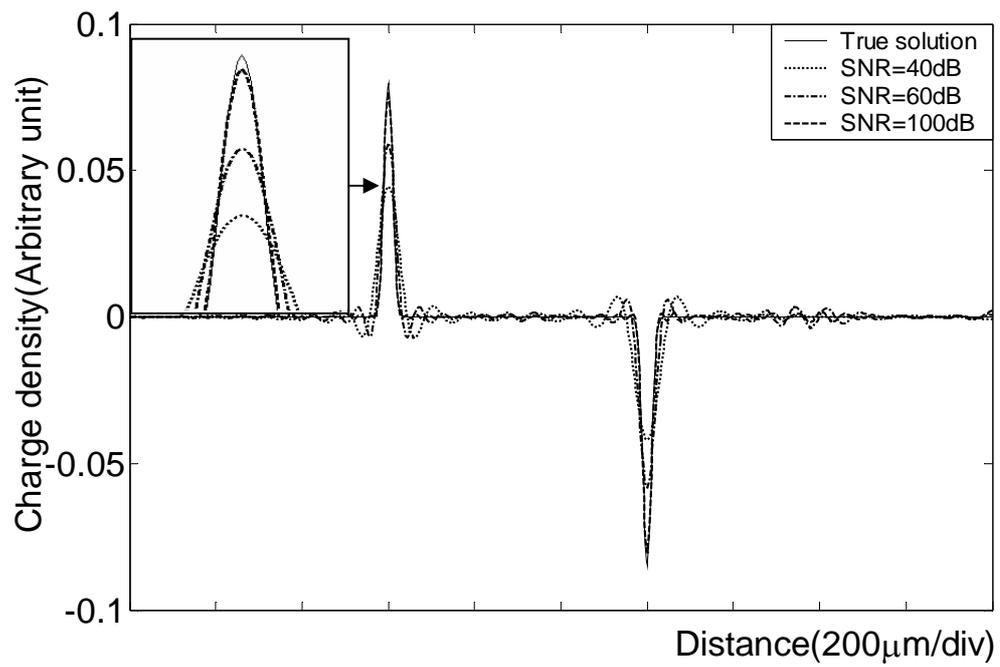

Figure 4. Luoquan HU *et al*

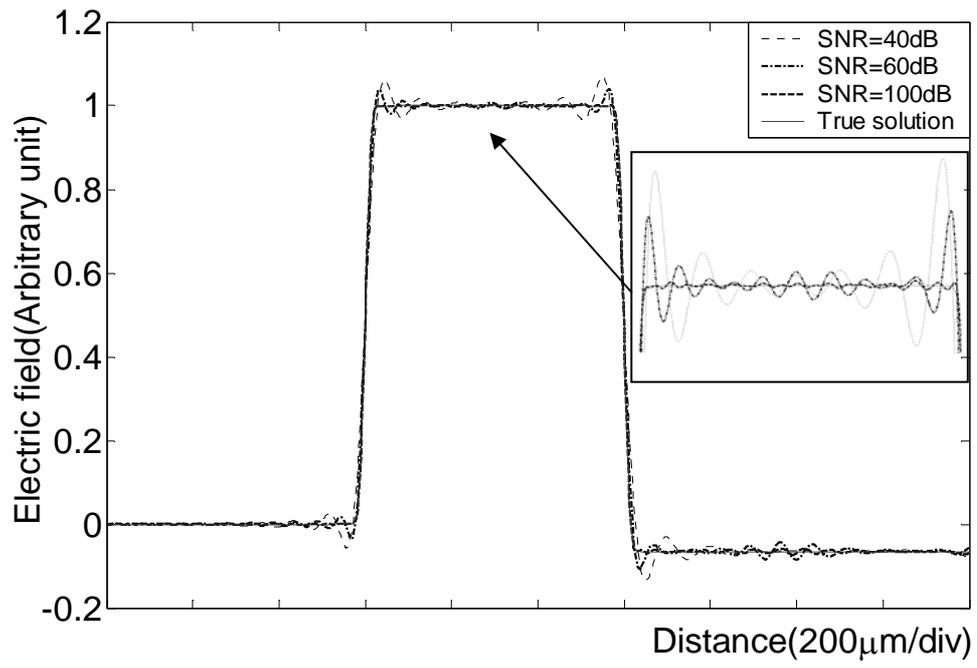

Figure 5. Luoquan HU *et al*

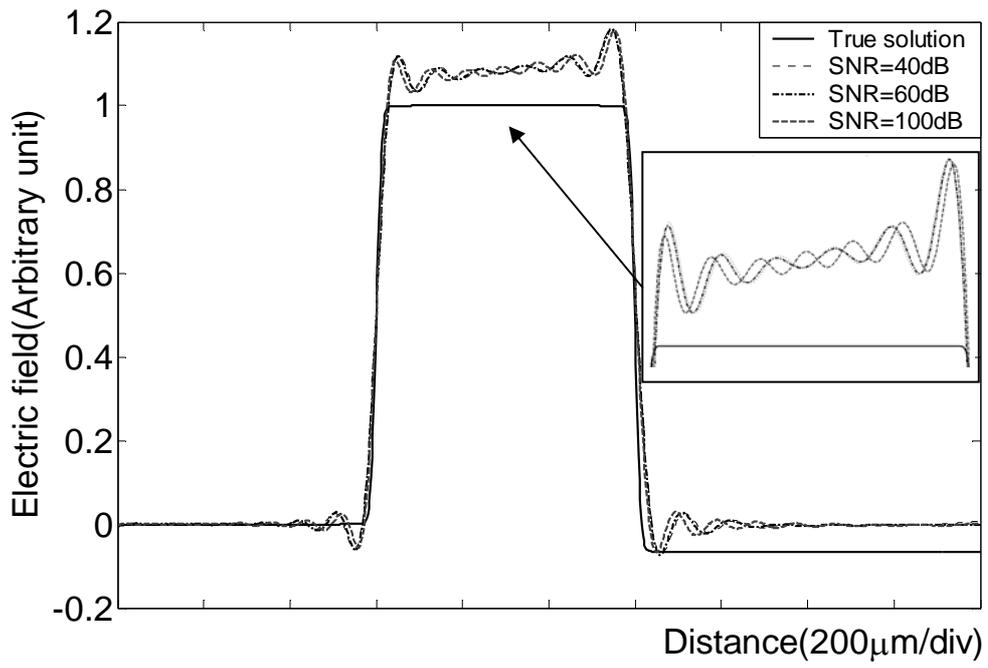

Figure 6. Luoquan HU *et al*

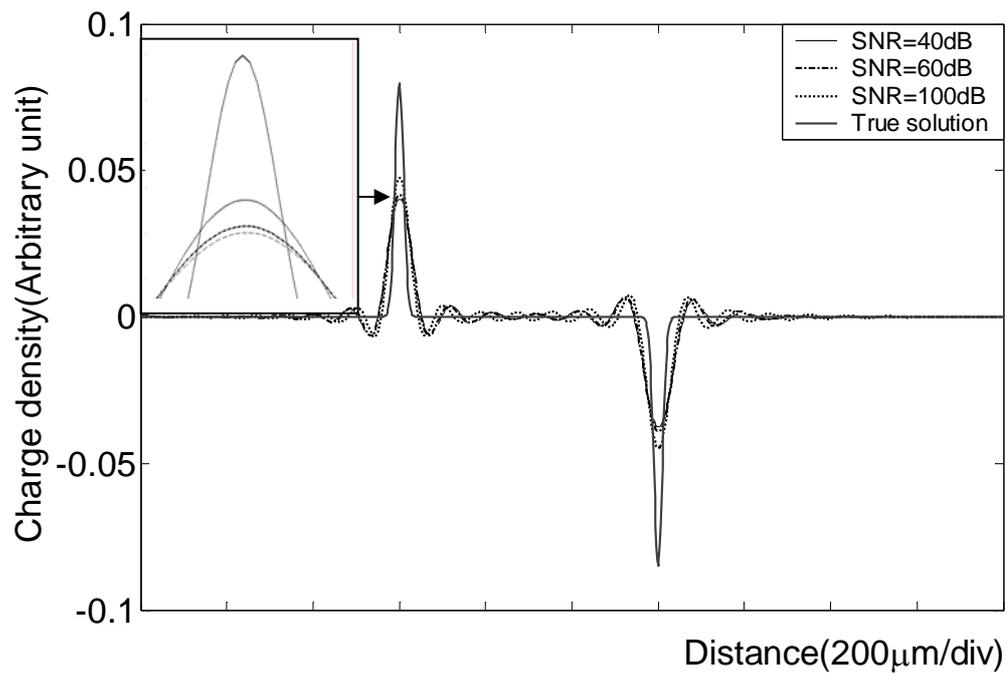

Figure 7. Luoquan HU *et al*

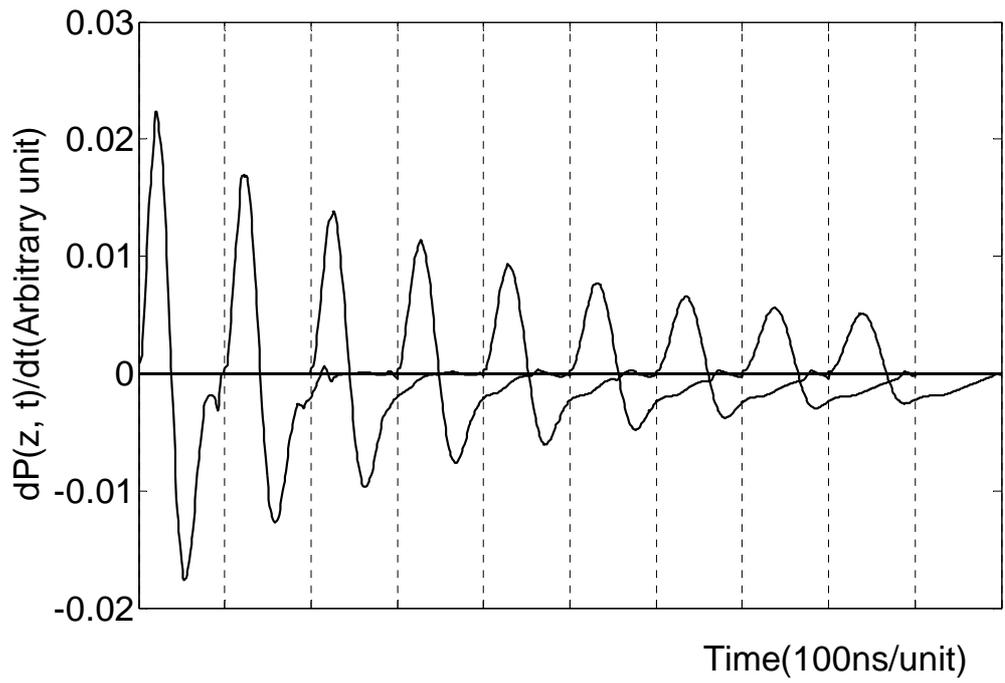

Figure 8. Luoquan HU *et al*

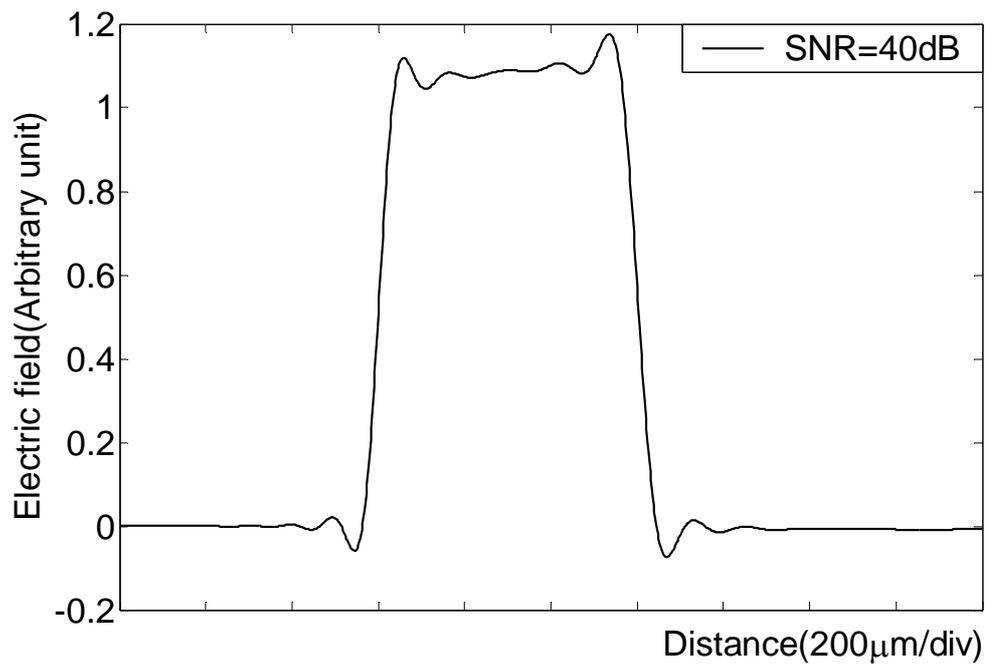

Figure 9. Luoquan HU *et al*

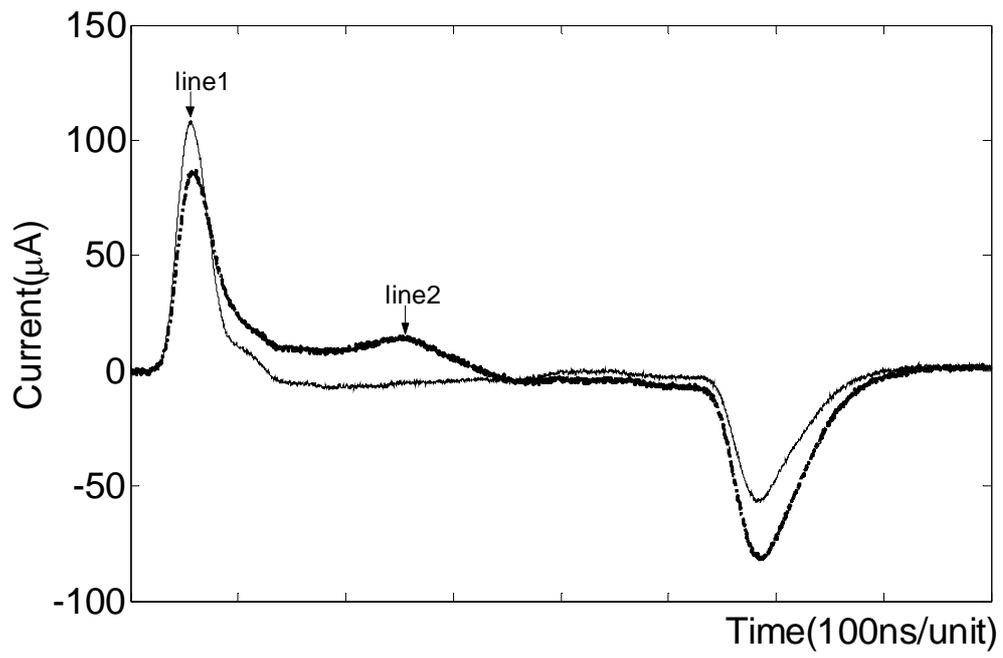

Figure 10. Luoquan HU *et al*

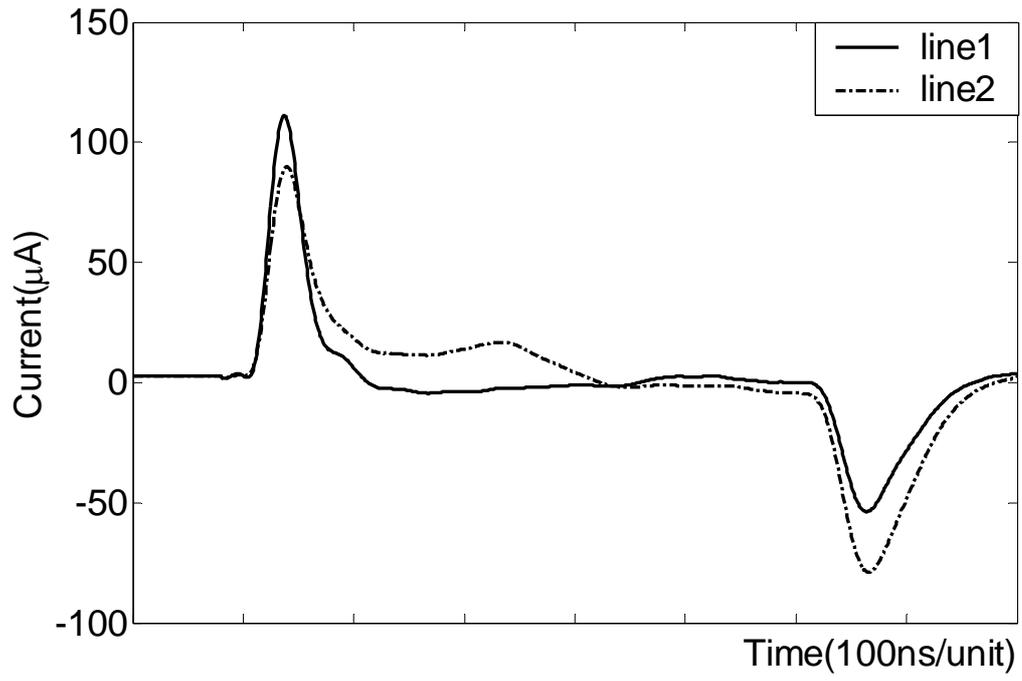

Figure 11. Luoquan HU *et al*

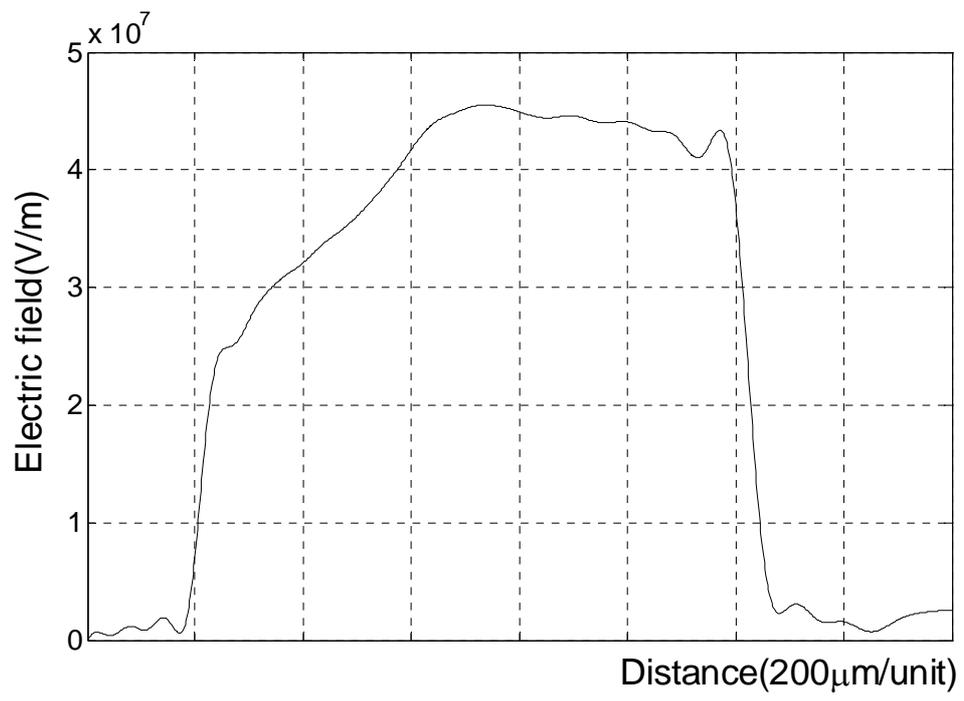

Figure 12. Luoquan HU *et al*